\documentclass[12pt]{article}
\usepackage{amssymb}
\usepackage{amsmath,epsfig}
\usepackage{graphicx,epsf,epsfig}
\usepackage{amsmath}
\usepackage{amsfonts}
\usepackage{mathrsfs}

\def\21{$SU(2) \otimes U(1)$}
\def\321{$SU(3) \otimes SU(2) \otimes U(1)$}

 \catcode`\@=11

\def\Re{\mbox{Re}\,}

\def\hbar{\hspace{0pt}\raisebox{1pt}{$-$} \hspace{-7pt} h}

\newcommand{\be}{\begin{equation}}
\newcommand{\ee}{\end{equation}}
\newcommand{\bd}{\begin{displaymath}}
\newcommand{\ed}{\end{displaymath}}
\newcommand{\bea}{\begin{eqnarray}}
\newcommand{\eea}{\end{eqnarray}}
\newcommand{\nn}{\nonumber}

\begin{document}

\title{Solving the SUSY Flavour and CP Problems with Non-Abelian Family
Symmetry and Supergravity}

\author{
 Stefan Antusch\footnote{\texttt{antusch@mppmu.mpg.de}}\;,\\[-1mm]  
{\small\em Max-Planck-Institut f\"ur Physik (Werner-Heisenberg-Institut), }\\[-2mm]
{\small\em F\"ohringer Ring 6, D-80805 M\"unchen, Germany,}\\[2mm]
 Stephen F.~King\footnote{\texttt{sfk@hep.phys.soton.ac.uk}}\;, Michal Malinsk\'{y}\footnote{\texttt{malinsky@phys.soton.ac.uk}}\;,\\[-1mm] 
{\small\em  School of Physics and Astronomy, University of Southampton,}\\[-2mm] 
{\small\em SO16 1BJ Southampton, United Kingdom,}\\[2mm]
Graham G.~Ross\footnote{\texttt{g.ross1@physics.ox.ac.uk}}\;,\\[-1mm]  
{\small\em The Rudolf Peierls Centre for Theoretical Physics,
The University of Oxford,
}\\[-2mm]
{\small\em 1 Keble Road, Oxford, OX13NP,
United Kingdom.}
}

\maketitle
\vspace{-0.2cm}
\begin{abstract}
Can a theory of flavour capable of describing the spectrum of fermion
(including neutrino) masses and mixings also contain within it the seeds for
a solution of the SUSY flavour and CP problems? We argue that supergravity
together with a non-Abelian family symmetry can completely resolve the SUSY
flavour and CP problems in a broad class of theories in which family 
symmetry and CP is spontaneously broken in the flavon sector. We show that a
simple superpotential structure can suppresses the $F$-terms of the 
flavons and GUT scale Higgs fields and that, if this mechanism is implemented, 
the resulting flavour and CP violation is suppressed and comfortably within 
the experimental limits. For illustration, we study a specific model based on 
$SU(3)$ family symmetry, but similar models based on non-Abelian (continuous 
or discrete) family symmetry will lead to similar results.
\end{abstract}

\vspace{1.4cm}

\section{Introduction}

The origin of the flavour structure of the quark and lepton masses and
mixing angles is one of the deepest mysteries left unanswered by the
Standard Model (SM) and remains one of the main motivations to go beyond it.
The introduction of Supersymmetry (SUSY), whilst providing plausible answers
to other mysteries left unanswered by the SM (such as the stability and
origin of the weak scale, the origin of dark matter, and the question of
unification), does not address the origin of flavour. In fact the
introduction of TeV scale SUSY gives rise to large flavour changing neutral
currents (FCNCs) and electric dipole moments (EDMs), larger than those
predicted by the SM, and potentially above the experimental limits 
\cite{Chung:2003fi}.

It has recently been demonstrated that $SU(3)$ family symmetry 
\cite{King:2001uz}, \cite{King:2003rf} can solve the flavour problem of the SM.
In this approach the smallness of neutrino masses is due to the see-saw
mechanism \cite{Minkowski:1977sc}, and the large lepton mixing is due to the
sequential dominance (SD) mechanism \cite{King:1998jw}, \cite{King:1999mb}.
Indeed present neutrino oscillation data is consistent with approximate
tri-bimaximal lepton mixing \cite{Harrison:2002er}, and this can be readily
achieved with constrained sequential dominance (CSD) \cite{King:2005bj}, 
\cite{deMedeirosVarzielas:2005ax}. In such family symmetry models a
potential solution to the SUSY\ CP\ problem results if the origin of CP
violation is due to the spontaneous breaking of the $SU(3)$ family symmetry
via flavon vacuum expectation values (vevs), $\langle \Phi _{i}\rangle $ 
\cite{Ross:2002mr}, \cite{Ross:2004qn}. In this case CP violation originates
in the flavour changing sector (where it is observed to be large) and CP
violation in the flavour conserving sector is suppressed by powers of small
mixing angles.

In a recent paper three of us \cite{Antusch:2007re} analysed this solution
of the SUSY flavour and CP problems in a model with gauged $SU(3)$ family
symmetry previously introduced to describe quark and lepton masses and
mixings, and in particular to generate neutrino tri-bimaximal mixing via
CSD. We performed a detailed bottom-up operator analysis of the soft SUSY
breaking Lagrangian in terms of a spontaneously broken $SU(3)$ family
symmetry, where the operator expansions are $SU(3)$-symmetric. We then made
a careful estimate of the mass insertion parameters describing flavour
changing and CP violation, keeping track explicitly of all the
coefficients, including a careful treatment of canonical normalization
effects. The results of this analysis showed that, while all the
experimental constraints coming from flavour changing and CP violation may
be evaded in such a framework, there remained a tension between theory and
experiment for $\mu \rightarrow e\gamma $ and the EDMs \cite{Antusch:2007re}. 
The common origin of both of these sources of tension resides in the soft
SUSY breaking trilinear couplings as was first pointed out for $SU(3)$
family symmetry models without in the framework of supergravity in \cite%
{Ross:2002mr}, \cite{Ross:2004qn}. This study was done in the context of
global SUSY and has recently been complemented by another study\ in the
context of supergravity (SUGRA) \cite{Calibbi:2008qt} of models related to
those of \cite{Ross:2004qn}. Taking the commonly assumed 
value for the $F$-terms of the flavons, $F_{\Phi
}\approx m_{3/2}\langle \Phi \rangle ,$ it was found that the experimental
bounds on $\mu \rightarrow e\gamma $ and the EDMs exclude significant
regions of parameter space and, in the allowed regions, are close to the
current experimental limits for SUSY\ states light enough to be produced at
the LHC.

In this paper we extend these studies to SUGRA$\ $models with an underlying $%
SU(3)$ family symmetry spontaneously broken in a manner that generates
tri-bimaximal mixing. With the commonly assumed value for the flavon $F$-terms we
again find a tension with the experimental bounds for $\mu \rightarrow
e\gamma $ and the EDMs. However we show that this tension can be removed via
a simple mechanism for suppressing the flavon $F$-terms below their 
commonly assumed value. Though initially formulated in the context of models with $SU(3)$ 
family symmetry, the mechanism does not rely on the particular choice of the 
flavour group and can be generalized to other scenarios.

The paper is organized as follows: In Section 2 we remind the reader about salient
features of the class of models with $SU(3)$ family symmetry, first in the 
globally supersymmetric context and then in SUGRA focusing on the potential 
tension with respect to the experimental data on lepton flavour violation and 
EDMs. We recapitulate the simple estimate for the commonly assumed values of the visible 
sector SUGRA $F$-terms and argue that supergravity alone does not provide a relief 
to the strain without further model-building. In Section 3 we discuss a dynamical 
mechanism yielding a further suppression of the visible sector $F$-terms 
which is capable of restoring the full compatibility of the $SU(3)$ model with 
experimental constraints. In Section 4 we conclude. Some of the technical details 
are deferred to a set of Appendices.

\section{$\boldsymbol{SU(3)}$ as an effective family symmetry}

We focus on the class of $SU(3)$ flavour models discussed in \cite%
{deMedeirosVarzielas:2005ax,deMedeirosVarzielas:2006fc} that describes the
observed quark and lepton masses and mixings, and in particular generates
neutrino tri-bimaximal mixing. These models require three specific types of
flavon fields, each of which is an anti-triplet of the $SU(3)$ family
symmetry, and each of which has a particular type of vacuum alignment,
namely: $\phi _{3}\sim (0,0,1)$, $\phi _{23}\sim (0,-1,1)$, $\phi _{123}\sim
(1,1,1)$, up to phases. In practice, the desired vacuum alignment must also
ensure that $\phi _{23}^{\dag }\phi _{123}=0$, in accordance with the CSD
requirements necessary to yield tri-bimaximal neutrino mixing \cite{King:2005bj}, \cite{deMedeirosVarzielas:2005ax}. The resulting form of the
effective Yukawa superpotential is 
\begin{equation}
\tilde{W}_{Y}=f_{a}f_{b}^{c}\frac{1}{M_{f}^{2}}\langle y_{1}^{f}(\phi
_{123})_{a}(\phi _{23})_{b}+y_{2}^{f}(\phi _{23})_{a}(\phi
_{123})_{b}+y_{3}^{f}(\phi _{3})_{a}(\phi _{3})_{b}+\frac{y_{\Sigma }^{f}}{%
M_{f}^{\Sigma }}(\phi _{23})_{a}(\phi _{23})_{b}\Sigma \rangle H
\label{YukawaSUGRA}
\end{equation}%
where we have written the left-handed fermions as $f_{a}$ and the
CP-conjugated right-handed fermions (both family triplets) as $f_{b}^{c}$,
where $f=u,d,e,\nu $ and $a,b=1,2,3$ are the $SU(3)$ family symmetry
indices. $H$ is the Higgs doublet superfield and the field $\Sigma $ is a
field whose vev generates a relative factor of 3 between the muon and
strange quark mass at the unification scale, implementing the Georgi
Jarlskog mechanism \cite{Georgi:1979df}, and providing a phenomenologically
appealing account of charged lepton and down quark masses. In addition the $%
\Sigma $ vev suppresses the contribution of the last term to the neutrinos 
\cite{Ross:2002fb}, allowing the remaining operators to give rise to
tri-bimaximal neutrino mixing via the CSD mechanism. The messenger mass, $%
M_{f},$ and the magnitude of flavon vevs are chosen to generate the desired
hierarchical form of the Dirac masses as discussed in detail in \cite%
{deMedeirosVarzielas:2005ax,deMedeirosVarzielas:2006fc} to which we refer
the reader for more details.

At the level of an effective theory (i.e. for energies well below $M_{Pl}$
in Planck-scale mediated SUSY breakdown in SUGRA) one can parametrize the
effective soft-SUSY breaking in terms of the coefficients of effective
operator expansions. For later convenience in comparing with the
experimental bounds we adopt the notation of \cite{Antusch:2007re}. Then the
leading order contributions have the form:  
\begin{eqnarray}
(\hat{m}_{f,f^{c}}^{2})_{\overline{a}b}&=& m_{0}^{2}\left(b_{0}^{f,f^{c}}\delta _{%
\overline{a}b}+b_{1}^{f,f^{c}}\frac{\langle \phi _{123}^{\ast }\rangle _{%
\overline{a}}\langle \phi _{123}\rangle _{b}}{M_{f}^{2}}+b_{2}^{f,f^{c}}%
\frac{\langle \phi _{23}^{\ast }\rangle _{\overline{a}}\langle \phi
_{23}\rangle _{b}}{M_{f}^{2}}+b_{3}^{f,f^{c}}\frac{\langle \phi _{3}^{\ast
}\rangle _{\overline{a}}\langle \phi _{3}\rangle _{b}}{M_{f}^{2}}\right.
\nonumber \\
&+&\left.b_{4}^{f,f^{c}}\delta _{\overline{a}b}\frac{\langle \Sigma ^{\ast
}\rangle \langle \Sigma \rangle _{{}}}{M_{\Sigma }^{2}}\right) 
\,,\\
\hat{A}_{ab}^{f}&=&A_{0}\frac{1}{M_{f}^{2}}\left( a_{1}^{f}\langle \phi
_{123}\rangle _{a}\langle \phi _{23}\rangle _{b}+a_{2}^{f}\langle \phi
_{23}\rangle _{a}\langle \phi _{123}\rangle _{b}+a_{3}^{f}\langle \phi
_{3}\rangle _{a}\langle \phi _{3}\rangle _{b}+\frac{a_{\Sigma }^{f}}{%
M_{f}^{\Sigma }}\langle \phi _{23}\rangle _{a}\langle \phi _{23}\rangle
_{b}\left\langle \Sigma \right\rangle \right),\nonumber
\end{eqnarray}%
where we have taken the messenger mass to be the same as that in the
expression for the Dirac mass matrices. At this point, the coefficients of
these expansions are generic numbers governing the phenomenology analysis.
However, once the SUSY breaking mechanism and messenger sector is specified,
these parameters become calculable.

\subsection{$\boldsymbol{SU(3)}$ family symmetry in Supergravity}

In addition to the fermion Dirac mass structure of Eq.~(\ref{YukawaSUGRA}) it
is necessary to specify the general form of the visible sector piece of the K%
\"{a}hler potential. In leading order it is parameterized by the term $%
\tilde{K}_{\overline{a}b}^{f}f^{\dagger \overline{a}}f^{b}+\tilde{K}_{%
\overline{a}b}^{f^{c}}f^{c\dagger \overline{a}}f^{cb}$ where%
\begin{eqnarray}
\tilde{K}_{\overline{a}b}^{f,f^{c}} &=&\delta _{\overline{a}b}\left(
k_{0}^{f,f^{c}}+l_{0}^{f,f^{c}}\frac{X^{\dagger }X}{M_{\mathrm{Pl}}^{2}}%
\right) +\frac{(\phi _{123}^{\ast })_{\overline{a}}(\phi _{123})_{b}}{%
M_{f}^{2}}\left( k_{1}^{f,f^{c}}+l_{1}^{f,f^{c}}\frac{X^{\dagger }X}{M_{%
\mathrm{Pl}}^{2}}\right)  \label{SU3Kahler} \\
&+&\frac{(\phi _{23}^{\ast })_{\overline{a}}(\phi _{23})_{b}}{M_{f}^{2}}%
\left( k_{2}^{f,f^{c}}+l_{2}^{f,f^{c}}\frac{X^{\dagger }X}{M_{\mathrm{Pl}%
}^{2}}\right) +\frac{(\phi _{3}^{\ast })_{\overline{a}}(\phi _{3})_{b}}{%
M_{f}^{2}}\left( k_{3}^{f,f^{c}}+l_{3}^{f,f^{c}}X^{\dagger }X\right) +\delta
_{\overline{a}b}k_{4}^{f,f^{c}}\frac{\Sigma ^{\dagger }\Sigma }{M_{\Sigma
}^{2}}\nn
\end{eqnarray}%
and $k_{i}^{ff^{c}},$ $l_{I}^{f,f^{c}}$are constants and $X$ denotes a
hidden sector field driving the SUSY breakdown. In this we have again
assumed that the messenger mass is the same as that in the expression for the
Dirac mass matrices.

In terms of Eqs.~(\ref{SU3Kahler}) and (\ref{YukawaSUGRA}) the generic
formulae for the leading order effective soft SUSY-breaking terms is%
\begin{equation}
\hat{m}_{\overline{a}b}^{2}\approx \left\langle m_{3/2}^{2}\tilde{K}_{\overline{a}%
b}-F_{X^{\dagger }}\left( \partial _{X^{\dagger }}\partial _{X}\tilde{K}_{%
\overline{a}b}\right) F_{X}-\sum_{\Phi _{I},\Phi _{J}}F_{{\Phi }_{I}^{\ast
}}\left( \partial _{{\Phi }_{I}^{\ast }}\partial _{\Phi _{J}}{\tilde{K}}_{%
\overline{a}b}^{f,f^{c}}\right) F_{\Phi _{J}}+\ldots \right\rangle 
\label{genericm}
\end{equation}%
and 
\begin{eqnarray}
A_{abc} &\propto &\left\langle F_{X}\left( \partial _{X}\frac{K_{\mathrm{hid}}}{%
M_{Pl}^{2}}\right) Y_{abc}+\sum_{\Phi }F_{\Phi }\partial _{\Phi }Y_{abc}
\label{generictrilinears} \right.\\
&-&\left.\left( F_{X}(\tilde{K}^{-1})_{d\overline{e}}\partial _{X}\tilde{K}_{%
\overline{e}a}Y_{dbc}+\sum_{\Phi }F_{\Phi }(\tilde{K}^{-1})_{d\overline{e}%
}\partial _{\Phi }\tilde{K}_{\overline{e}a}Y_{dbc}+cyclic{(a,b,c)}\right)
\right\rangle\,,   \notag
\end{eqnarray}%
where $\Phi $ stands for all the visible sector fields in the model (in
particular the flavons $\phi _{3}$, $\phi _{23},$ $\phi _{123}$ and also $%
\Sigma $).

In the formulae above we have assumed, c.f.\ Eq.~(\ref{YukawaSUGRA}), that due
to the holomorphy of the superpotential the direct couplings of $X$ to the
Yukawa sector of the model are absent (which leads to the absence of the $%
\partial _{X}Y_{abc}$ terms in Eq.~(\ref{generictrilinears})). This is
the case for the specific family symmetry model discussed above.

\subsection{Commonly assumed value for SUGRA $\boldsymbol{F}$-terms\label{sect-genericF}}

It is obvious from Eqs.~(\ref{genericm}) and (\ref{generictrilinears}) that
the values of the various $F$-terms provide a crucial ingredient of any
detailed analysis of soft terms. In SUGRA the ``natural'' expectation for $F-$%
terms of visible sector superfields, $\Phi ,$ whose scalar component
acquires a vev is given by $F_{\Phi }\approx m_{3/2}\langle \Phi \rangle $,
up to cancellations \cite{Ross:2002mr}. This stems from the fact that in
Planck units the generic structure of SUGRA $F$-terms is given by 
\begin{equation}
F_{I}=-\,e^{G/2}(G^{-1})_{I\overline{J}}G_{\overline{J}}=-%
\,e^{G/2}(K^{-1})_{I\overline{J}}G_{\overline{J}}\, , 
\end{equation}%
where 
\begin{equation*}
G_{\overline{J}}\equiv \partial _{\overline{J}}\left( {K}+\log {W^{\ast }}%
+\log {W}\right) =(W^{\ast })^{-1}\left( W^{\ast }{K}_{\overline{J}}+W_{%
\overline{J}}^{\ast }\right)
\end{equation*}
and thus 
\begin{equation*}
F_{I}=-e^{K/{2}}(K^{-1})_{I\overline{J}}\left( W^{\ast }{K}_{\overline{J}%
}+W_{\overline{J}}^{\ast }\right)\,.
\end{equation*}
In the $M_{Pl}\rightarrow \infty $ limit only the global SUSY term $%
F_{I}\propto -(K^{-1})_{I\overline{J}}W_{\overline{J}}^{\ast }$ survives.
Plugging in the gravitino mass $m_{3/2}^{2}=e^{\langle K\rangle }\langle {%
|W|^{2}}\rangle $ one arrives at 
\begin{equation}
F_{I}=-(K^{-1})_{I\overline{J}}\left( m_{3/2}{K}_{\overline{J}}-e^{K/{2}}W_{%
\overline{J}}^{\ast }\right)\,.  \label{Ftermshape}
\end{equation}%
For the standard K\"{a}hler potential $\Phi \Phi ^{\ast }$ the first term in
(\ref{Ftermshape}) provides an \textquotedblleft
irreducible\textquotedblright\ contribution to the relevant $F$-term given
by 
\begin{equation}
\left\langle F_{\Phi }\right\rangle =m_{3/2}\langle \Phi \rangle
\label{naturalF}\,.
\end{equation}%
Unless there are 
cancellations from the second ($W_{\overline{J}%
}^{\ast }$) term in Eq.~(\ref{Ftermshape}) this provides a lower bound on the 
$F$-term. Taking this as a starting point we parameterize the $F$-terms of
the visible sector superfields in the model by

\begin{equation}
F_{(\phi _{A})_{i}}\equiv m_{3/2}x_{A}\left\langle \phi _{A}\right\rangle
_{i}+\ldots ,\qquad F_{\Sigma }\equiv m_{3/2}x_{\Sigma }\left\langle \Sigma
\right\rangle +\ldots  \label{genericFterms}
\end{equation}%
and give all our leading order results in terms of the $x_{A}$ and $x_{\Sigma }$ 
factors (note that the gauge and family symmetries of the model
ensure that at the leading order the $F$-terms are diagonal in the field
space).

\subsection{The soft SUSY-breaking terms}

Using Eq.~(\ref{genericFterms}) in Eq.~(\ref{genericm}) the soft masses are
given by 
\begin{eqnarray}
(\hat{m}_{f,f^{c}}^{2})_{\overline{a}b} &=&m_{3/2}^{2}\left[\delta _{\overline{a}%
b}\left( k_{0}^{f,f^{c}}+l_{0}^{f,f^{c}}\frac{\left\langle X^{\dagger
}X\right\rangle }{M_{\mathrm{Pl}}^{2}}-l_{0}^{f,f^{c}}\frac{F_{X^{\dagger
}}F_{X}}{m_{3/2}^{2}M_{Pl}^{2}}\right)  \label{softmassescomputed}\right. \\
&+&\left.\sum_{A}\frac{\left\langle \phi _{A}^{\ast }\right\rangle _{\overline{a}%
}\left\langle \phi _{A}\right\rangle _{b}}{M_{f}^{2}}\left(
k_{A}^{f,f^{c}}\left( 1-x_{A}x_{A}^{\ast }\right) -l_{A}^{f,f^{c}}\frac{%
F_{X^{\dagger }}F_{X}}{m_{3/2}^{2}M_{Pl}^{2}}\right) -k_{4}^{f,f^{c}}\delta
_{\overline{a}b}\frac{\left\langle \Sigma ^{\ast }\right\rangle \left\langle
\Sigma \right\rangle }{M_{\Sigma }^{2}}x_{\Sigma }x_{\Sigma }\right]\,.\nonumber
\end{eqnarray}%
The relevant dictionary between the SUGRA setting and the operator
coefficients relevant for the effective analysis reads: $m_{0}=m_{3/2}$,
\begin{equation}
b_{0}^{f,f^{c}}=k_{0}^{f,f^{c}}+l_{0}^{f,f^{c}}\delta _{%
\overline{a}b}\left( \frac{\left\langle X^{\dagger }X\right\rangle }{{%
M_{Pl}^{2}}}-\frac{F_{X^{\dagger }}F_{X}}{m_{3/2}^{2}{M_{Pl}^{2}}}\right)
,\;\;b_{A}^{f,f^{c}}=k_{A}^{f,f^{c}}\left( 1-x_{A}x_{A}^{\ast }\right)
-l_{A}^{f,f^{c}}\frac{F_{X^{\dagger }}F_{X}}{M_{Pl}^{2}}.
\end{equation}

Turning to the trilinear terms, as we now discuss, the dominant term is the
second one in Eq.~(\ref{generictrilinears}). Consider first the flavour
violating terms. Since the first term in Eq.~(\ref{generictrilinears})
is proportional to the relevant Yukawa matrix it does not contribute to
flavour violation. Next, due to the $\partial _{\phi }\tilde{K}_{\overline{a}%
b}$ factor, the terms proportional to $F_{\Phi }$ in the square bracket in
Eq.~(\ref{generictrilinears}) are at least two powers of $\left\langle \phi
\right\rangle /M^{f}$ more suppressed than the second term. Concerning the
terms coming from the first and last terms of Eq.~(\ref{SU3Kahler}), they are
proportional to the Yukawa matrix and hence are also flavour conserving.
Finally the terms proportional to $F_{X}$ coming from the second, third and
fourth terms of Eq.~(\ref{SU3Kahler}) are at least two powers of 
$\left\langle \phi \right\rangle /M^{f}$ more suppressed than the leading
terms coming from the second term of Eq.~(\ref{generictrilinears}). Thus, the 
$\left\langle F_{\Phi }\partial _{\Phi }Y_{abc}\right\rangle $ terms govern
the SUSY flavour-violation in the trilinear couplings. On the CP side we
shall focus on the (flavour conserving) EDMs that provide the most
stringent bounds on the CP phases in the trilinear sector. Recall that the
relevant quantities are the (1,1)-entries of the trilinear couplings in the
super-CKM basis. As shown in \cite{Ross:2002mr} the imaginary parts on the
diagonal of the SCKM-basis trilinears from the terms in the square bracket
in Eq.~(\ref{generictrilinears}) are suppressed. Assuming the hidden sector
fields $X$ do not couple to the Yukawa sector and the $F_{\Phi }$-terms have
their commonly assumed values, the leading
contributions to the SUSY EDMs also come predominantly from the $F_{\Phi
}\partial _{\Phi }Y_{abc}$ terms in Eq.~(\ref{generictrilinears}).

From Eqs.~(\ref{YukawaSUGRA}) and (\ref{genericFterms}) we have 
\begin{eqnarray}
\left\langle F_{\Phi }\partial _{\Phi }{Y}^{f}\right\rangle _{ab}&\approx&
\langle \Bigl[\sum_{A,c}F_{(\phi _{A})_{c}}\partial _{(\phi
_{A})_{c}}+F_{\Sigma }\partial _{\Sigma }\Bigr]\hat{Y}_{ab}^{f}\rangle
\label{SU3Aterms} \\
&=&\frac{m_{3/2}}{M_{f}^{2}}\Big\langle\left[ y_{1}^{f}(\phi _{123})_{a}(\phi
_{23})_{b}+y_{2}^{f}(\phi _{23})_{a}(\phi _{123})_{b}\right] \left(
x_{23}+x_{123}\right)+2y_{3}^{f}(\phi _{3})_{a}(\phi _{3})_{b}x_{3} \nonumber\\
&+&\frac{y_{\Sigma }^{f}}{%
M_{f}^{\Sigma }}(\phi _{23})_{a}(\phi _{23})_{b}\Sigma \left(
2x_{23}+x_{\Sigma }\right) \Big\rangle\nonumber
\end{eqnarray}%
which, c.f.\ Eq.~(\ref{generictrilinears}), gives:%
\begin{equation}
a_{1}^{f}=y_{1}^{f}\left( x_{123}+x_{23}\right)
,\;\;a_{2}^{f}=y_{2}^{f}\left( x_{123}+x_{23}\right)
,\;a_{3}^{f}=y_{2}^{f}\left( 2x_{3}\right) ,\;a_{\Sigma }^{f}=y_{\Sigma
}^{f}\left( 2x_{23}+x_{\Sigma }\right) ,\;\;A_{0}=m_{3/2}.
\label{dictionaryA}
\end{equation}

\subsection{Phenomenology of the SUGRA $\boldsymbol{SU(3)}$ model}

\subsubsection{Lepton flavour violation ($\boldsymbol{\protect\mu \rightarrow e \gamma}$)}

It is straightforward now to use the analysis of \cite{Antusch:2007re} to
determine the phenomenological implications of the $SU(3)$ model. Using the
coefficients just determined, the relevant mass insertion parameter, $\delta$, 
governing the branching ratio of $\mu \rightarrow e\gamma $ gives 
\begin{equation}
|(\delta _{LR}^{\ell })_{12}|\approx 1\times 10^{-4}{\frac{{A_{0}}}{{100\,%
\mathrm{GeV}}}}\frac{(200\,\mathrm{\ GeV})^{2}}{\langle \tilde{m}_{l}\rangle
_{LR}^{2}}\frac{10}{\tan \beta }\left( \frac{\overline{\varepsilon }}{0.13}%
\right) ^{3}|y_{1}|\left\vert x_{123}-x_{23}-x_{\Sigma }\right\vert\,.
\label{globalmutoegamma}
\end{equation}

\subsubsection{Electric dipole moments}

Similarly, the mass insertion parameters determining the EDMs are 
\begin{eqnarray}
\!\!|\mathrm{Im}(\delta _{LR}^{u})_{11}|\!\! &\!\!\approx \!\!&\!2\!\times \!10^{-7}%
\frac{A_{0}}{100\text{\,GeV}}\!\!\left( \frac{500\,\mathrm{GeV}}{\langle 
\tilde{m}_{u}\rangle _{LR}}\right) ^{2}\!\!\left( \frac{\overline{%
\varepsilon }}{0.13}\right) ^{3}\!\!\!\left( \frac{\varepsilon }{0.05}%
\right) ^{2}
|y_{1}^{f}+y_{2}^{f}|\left\vert
x_{123}-x_{23}-x_{\Sigma }\right\vert
\sin \phi _{1}  \notag, \\
\!\!|\mathrm{Im}(\delta _{LR}^{d})_{11}|\!\! &\!\!\approx \!\!&\!5\!\times \!10^{-7}%
\frac{A_{0}}{100\text{\,GeV}}\!\!\left( \frac{500\,\mathrm{GeV}}{\langle 
\tilde{m}_{d}\rangle _{LR}}\right) ^{2}\!\!\left( \frac{\overline{%
\varepsilon }}{0.13}\right) ^{5}\!\!\frac{10}{\tan \beta }
|y_{1}^{f}+y_{2}^{f}|\left\vert
x_{123}-x_{23}-x_{\Sigma }\right\vert
\sin \phi _{1},  \label{EDMs} \\
\!\!|\mathrm{Im}(\delta _{LR}^{\ell })_{11}|\!\! &\!\!\approx \!\!&\!2\!\times \!10^{-7}%
\frac{A_{0}}{100\text{\,GeV}}\!\!\left( \frac{200\,\mathrm{GeV}}{\langle 
\tilde{m}_{e}\rangle _{LR}}\right) ^{2}\!\!\left( \frac{\overline{%
\varepsilon }}{0.13}\right) ^{5}\!\!\frac{10}{\tan \beta }
|y_{1}^{f}+y_{2}^{f}|\left\vert
x_{123}-x_{23}-x_{\Sigma }\right\vert
\sin \phi _{1},  \notag
\end{eqnarray}%
where $\phi _{1}$ is a CP phase associated to the VEV of the $\phi _{123}$
flavon and using Eq.~(\ref{dictionaryA}). We have chosen to normalize
the expansion parameters $\varepsilon $ and $\overline{\varepsilon }$ to the
values found in a recent fit to the measured masses and mixing angles 
\cite{Ross:2007az}. 

Thus, both $\mu \rightarrow e\gamma $ and the EDMs are determined by a
single combination, $\Delta ,$ of the $x$-factors which parametrize the
structure of the relevant visible sector $F$-terms, where 
\begin{equation}
\Delta \equiv \left\vert x_{123}-x_{23}-x_{\Sigma }\right\vert.
\label{bracket}
\end{equation}%
The present experimental bound from the non-observation of $\mu \rightarrow
e\gamma $ is $|(\delta _{LR}^{\ell })_{12}|\leq 10^{-5}$ which is in some
tension with this bound requiring, for example, $\tilde{m}_{l}=$ $600$\,GeV if
the remaining factors in Eq.~(\ref{globalmutoegamma}) are of $O(1)$. For the
EDMs the most stringent bound comes from mercury and corresponds to $|\mathrm{%
Im}(\delta _{LR}^{d})_{11}|<6.7\times 10^{-8}$ and requires $\tilde{m}%
_{d}=1500$\,GeV if the other factors are of $O(1)$. This means that SUGRA
does not automatically provide a relief from the flavour and CP issues of
the $SU(3)$ model under consideration as compared to the effective operator
analysis \cite{Antusch:2007re}, c.f.\ also \cite{Malinsky:2007pf}.

\section{Suppressing EDMs and $\boldsymbol{\protect\mu\to e \protect\gamma}$}

Given this tension it is appropriate to review the possibilities for
reducing $\Delta .$ The ``natural'' expectation is that $x_{i}\simeq 1,$
corresponding to the case $\left\langle \partial W/\partial \phi
_{i}\right\rangle \simeq 0.$ Although $\Delta $ does not vanish in this case
it is relatively easy to modify the model to arrange for it to do so. This
requires that each term in the mass matrix should involve the same number of
flavon and $\Sigma $ fields. A simple illustration of this mechanism is
given in the Appendix \ref{dummyfields}. Although this mechanism does work
it represents an unwanted complication of the model so here we concentrate
on a more promising possibility. It turns out that it is relatively easy to
modify the model to arrange for a non-zero $\left\langle \partial W/\partial
\phi _{i}\right\rangle $ to cancel $F_{\Phi }$ giving $x_{i}\simeq 0.$

\subsection{Dynamical suppression of $\boldsymbol{F}$-terms in SUGRA\label%
{genericmechanism}}

We begin by studying a simple class of models provided the following
conditions apply:

\begin{enumerate}
\item The K\"{a}hler potentials are all of the canonical form; since the
flavon vevs must be below the Planck scale this is a reasonable assumption
as the canonical form is the leading term allowed in the K\"{a}hler
potential in a power series expansion of the superfields.

\item The superpotential of the world can be written as $W=W_{obs}+W_{hid}$
where the superpotential of the observable sector may be written as $%
W_{obs}=Z(\Psi \overline{\Psi }-M_{\Psi }^{2})$ where $Z$ is some visible
sector superfield and the mass scale is below the Planck mass $M_{\Psi
}<M_{P}$ (notice that this form of $W_{obs}$ is the one which is typically
used in the globally-supersymmetric flavour models for sake of arranging the
desired patterns of flavon vevs);

\item Negligible $D$-terms. In fact this proves to be the case for a variety
of models. Although our discussion has concentrated on the case of a
continuous family symmetry, it applies equally to the case that the
structure of the superpotential and K\"{a}hler potential is driven by a
discrete non-Abelian subgroup of $SU(3)$ \cite{deMedeirosVarzielas:2006fc}. 
In this case there are no $D$-terms. Even in the case of a continuous $SU(3)$
the vacuum structure just discussed is not disturbed by $D$-terms as we
demonstrate in Appendix B. Of course one must also check that the $D$-terms
do not introduce unacceptable FCNCs or CP violation; that this is the case
is discussed in \cite{de Medeiros Varzielas:2006ma}.
\end{enumerate}
Under the above assumptions the scalar potential is given by 
\begin{equation}
V=\left( |F_{obs}|^{2}+|F_{hid}|^{2}- 3\, e^{K} |W|^{2}\right)\,,  \label{V}
\end{equation}%
where 
\begin{equation}
|F_{obs}|^{2}={\ |F_{\Psi }|^{2}+|F_{\overline{\Psi }}|^{2}+|F_{Z}|^{2}}
\label{VFvis}
\end{equation}%
and the individual observable sector $F$-terms may be written as 
\begin{equation}
F_{\Psi } \,\approx\, Z\Psi +m_{3/2}\overline{\Psi }^{\ast },\qquad F_{\overline{\Psi }%
} \,\approx\, Z\overline{\Psi }+m_{3/2}\Psi ^{\ast },\qquad F_{Z} \,\approx\, \Psi \overline{\Psi }%
-M_{\Psi }^{2}+m_{3/2}Z^{\ast }\,,  \label{Fterms}
\end{equation}%
where we have exploited the canonical form of the K\"{a}hler potential 
and used\footnote{%
Note that this result follows from the assumed forms $W=W_{obs}+W_{hid}$
where $W_{obs}=Z(\Psi \overline{\Psi }-M_{\Psi }^{2})$ which implies that at
the minimum of the potential $W_{obs}\ll W_{hid}$ which is plausible, given
the assumed form of $W_{obs}$, but which can be checked \textit{a posteriori}.} 
\begin{equation}
\left\langle |W_{hid}|\right\rangle \approx \left\langle |W|\right\rangle
\approx m_{3/2}\;,  \label{post}
\end{equation}
sticking to the leading contribution from the exponential.
We now argue that the potential is minimized for values of the visible
sector fields $Z,\Psi ,\overline{\Psi }$ such that $F_{\Psi }\ll
m_{3/2}\left\langle \Psi \right\rangle $ and $F_{\overline{\Psi }}\ll
m_{3/2}\left\langle \overline{\Psi }\right\rangle $. It is important to
emphasize that we are seeking a minimum of the potential in terms of the
observable fields $\Psi ,\overline{\Psi },Z$ and so we may expand the
potential as follows: 
\begin{equation}
V(\Psi ,\overline{\Psi },Z)\approx |F_{obs}|^{2}-3\left\vert
W_{obs}^{{}}\right\vert ^{2}-6\Re\left[ W_{hid}W_{obs}\right] +C\,,
\label{Vvis}
\end{equation}%
where only the leading order contribution of the exponential in Eq.~(\ref{V})
has been retained and $C$ is a constant term driven by the hidden sector
dynamics to account for a zero (or negligible) cosmological constant. A
necessary condition for a minimum of the potential is that the first
derivatives vanish ${\partial V}/{\partial Z}=0$, ${\partial V}/{%
\partial \Psi }=0$, ${\partial V}/{\partial \overline{\Psi }}=0$. By
explicit calculation it can readily be seen that the first derivatives
vanish for\footnote{%
Here we implicitly used Eq.~(\ref{post}) to give the result in a simple form. An
analytic minimization of the potential of Eq.~(\ref{Vvis}) yields: $%
|\left\langle \Psi \right\rangle \left\langle \overline{\Psi }\right\rangle
|=M_{\Psi }^{2}+m_{3/2}^{2}-3m_{3/2}\left\langle W_{hid}\right\rangle $ and $%
\left\langle Z\right\rangle =-m_{3/2}+3m_{3/2}^{2}\left\langle
W_{hid}\right\rangle /M_{\Psi }^{2}-9m_{3/2}\left\langle
W_{hid}\right\rangle ^{2}/2M_{\Psi }^{2}$ up to higher order terms and an 
overall phase (due to the phase difference between $\left\langle
\Psi \right\rangle $ and $\left\langle \overline{\Psi }\right\rangle $) in $%
\left\langle Z\right\rangle $.}: 
\begin{equation}
\left\langle Z\right\rangle = - m_{3/2}+O(m_{3/2}^{3}/M_{\Psi }^{2})\;,
\label{vac1}
\end{equation}%
\begin{equation}
\qquad |\left\langle \Psi \right\rangle |=|\left\langle \overline{\Psi }%
\right\rangle |\;\quad \text{and}\quad |\left\langle \Psi \right\rangle
\left\langle \overline{\Psi }\right\rangle |=M_{\Psi }^{2}+O(m_{3/2})^{2}
\label{vac2}
\end{equation}%
with anti-aligned phases on the components of $\left\langle \Psi
\right\rangle $ and $\left\langle \overline{\Psi }\right\rangle $ (up to a
possible global phase difference due to a would-be non-zero phase of $%
\left\langle Z\right\rangle $). Inserting these vevs into the $F$-terms in
Eq.~(\ref{Fterms}) it can be seen that 
\begin{equation}
|F_{\Psi }|=|F_{\overline{\Psi }}|= O(m_{3/2}^{2}/M_{\Psi }^{2})\times
m_{3/2}M_{\Psi }  \label{Fterms2}\,,
\end{equation}%
which is of the form $F_{\Psi }=x_{\Psi }m_{3/2}\left\langle \Psi
\right\rangle $ with a suppression factor of $x_{\Psi
}=O(m_{3/2}^{2}/M_{\Psi }^{2})$. Note that $\left\langle F_{Z}\right\rangle $
remains at its commonly assumed value $%
m_{3/2}\left\langle Z\right\rangle$. Moreover, at the minimum corresponding 
to the field values in Eqs.~(\ref{vac1})-(\ref{vac2}) we have $W_{obs}=O(m_{3/2}^{3})$ 
which justifies Eq.~(\ref{post}) \textit{a posteriori}.

It is straightforward to check that the configuration of Eqs.~(\ref{vac1})-(\ref{vac2}) 
corresponds to a (local) minimum by moving away slightly from the
minimum. In this case the variation is dominated by the
(non-Planck-suppressed) first term in Eq.~(\ref{V}) and clearly increases 
away from the turning point.

\subsection{EDMs and $\boldsymbol{\protect\mu\to e \protect\gamma}$ in SUGRA $\boldsymbol{SU(3)}$ with
dynamically \\ suppressed flavon and $\boldsymbol{\Sigma}$-field $\boldsymbol{F}$-terms}

The previous Section shows how $F$-terms can be suppressed below their
commonly assumed values. In Appendix B we
discuss how this can apply to one or more of the flavon fields and to the $%
\Sigma $ field. How can this suppression affect $\mu \rightarrow e\gamma $
and the EDMs? A particularly simple case is when the $\Sigma $ field has a
suppressed $F$-term while the flavons have their commonly assumed values. This makes $%
\Delta $ small due to the cancellation between the first two terms in Eq.~(%
\ref{bracket}). However, in this case, the cancellation is spoilt by the
next-to-leading contributions in Eq.~(\ref{YukawaSUGRA}) which introduce
different powers of the flavon fields in different matrix elements of the
Yukawa matrix. As a result our previous estimates are reduced by an extra
power of $\varepsilon $ in $|\mathrm{Im}(\delta _{LR}^{u})_{11}|$ and of $%
\overline{\varepsilon }$ in $|\mathrm{Im}(\delta _{LR}^{d})_{11}|$ and $|%
\mathrm{Im}(\delta _{LR}^{l})_{11}|$ respectively (for further details c.f.\ 
\cite{Antusch:2007re}) to get: 
\begin{eqnarray}
|\mathrm{Im}(\delta _{LR}^{u})_{11}| &\approx &1\times 10^{-8}\frac{A_{0}}{%
100\text{ GeV}}\left( \frac{500\,\mathrm{GeV}}{\langle \tilde{m}_{u}\rangle
_{LR}}\right) ^{2}\left( \frac{\overline{\varepsilon }}{0.13}\right)
^{3}\left( \frac{\varepsilon }{0.05}\right) ^{3}\sin \phi _{1}\,,
\label{SUGRAEDMsdummy} \notag\\
|\mathrm{Im}(\delta _{LR}^{d})_{11}| &\approx &6\times 10^{-8}\frac{A_{0}}{%
100\text{ GeV}}\left( \frac{500\,\mathrm{GeV}}{\langle \tilde{m}_{d}\rangle
_{LR}}\right) ^{2}\left( \frac{\overline{\varepsilon }}{0.13}\right) ^{6}%
\frac{10}{\tan \beta }\sin \phi _{1}\,,   \\
|\mathrm{Im}(\delta _{LR}^{\ell })_{11}| &\approx &3\times 10^{-8}\frac{A_{0}%
}{100\text{ GeV}}\left( \frac{200\,\mathrm{GeV}}{\langle \tilde{m}%
_{e}\rangle _{LR}}\right) ^{2}\left( \frac{\overline{\varepsilon }}{0.13}%
\right) ^{6}\frac{10}{\tan \beta }\sin \phi _{1}\,.  \notag
\end{eqnarray}%
As before, we are assuming $m_{0}(M_{\mathrm{GUT}})\approx 100$ GeV and take
into account the RG effects. These numbers are well below the current
experimental limits for all the elementary particle EDMs and compatible with
those on mercury EDM.

Similarly, for $\mu \rightarrow e\gamma $ one gets (multiplying the global
SUSY estimate of Eq.~(\ref{globalmutoegamma}) by an extra $\overline{\varepsilon }$%
): 
\begin{equation}
|(\delta _{LR}^{e})_{12}|\lesssim |(\delta _{LR}^{\ell })_{12}|\approx
1\times 10^{-5}{\frac{{A_{0}}}{{100\,\mathrm{GeV}}}}\frac{(200\,\mathrm{\ GeV%
})^{2}}{\langle \tilde{m}_{l}\rangle _{LR}^{2}}\frac{10}{\tan \beta }\left( 
\frac{\overline{\varepsilon }}{0.13}\right) ^{4}\,,
\label{SUGRAdummymutoegamma}
\end{equation}%
which is also compatible with the current bounds, in particular for large $%
\tan \beta $. 

The second possibility is that the flavons also have suppressed $F$-terms
corresponding to a suppression of $\Delta $ compared to its
commonly assumed value is the factor $%
O(m_{3/2}^{2}/M_{\Psi }^{2})$ of Eq.~(\ref{Fterms2}). In this case the
dominant contribution will be the term proportional to $F_{X}$ in the last
term in brackets in Eq.~(\ref{generictrilinears}), suppressed by two powers
of $\left\langle \phi \right\rangle /M^{f}$, corresponding to a further
suppression by the factor $\varepsilon $ in $|\mathrm{Im}(\delta
_{LR}^{u})_{11}|$ and of $\overline{\varepsilon }$ in $|\mathrm{Im}(\delta
_{LR}^{d})_{11}|$ and $|\mathrm{Im}(\delta _{LR}^{l})_{11}|$ respectively
taking the prediction well below the experimental limit even for the EDM of
mercury.

\section{Conclusions}

In this paper we have analysed the expectation for flavour changing neutral
currents and CP violating electric dipole moments in a class of
supergravity models with a non-Abelian family symmetry. With the 
commonly assumed values for the $F$-terms of the flavon and Georgi 
Jarlskog fields there is
tension between the experimental limits and the predicted values that
requires rather large SUSY particle masses. However we have identified a
simple mechanism for suppressing the $F$-terms and the resulting soft SUSY
breaking trilinear couplings. 
As a result the expectation for lepton flavour violating processes and EDMs
in these classes of models may be suppressed to values comfortably within
current limits. We emphasize again that the suppression mechanism presented
here is applicable to a very large class of models based on non-Abelian
(discrete or continuous) family symmetry and SUGRA, which henceforth should
be regarded as viable candidates for solving the SUSY flavour and CP
problems.
On the other hand even with the maximum suppression lepton number violating processes  
and EDMs are within a factor of 10 of present limits so future  
measurements capable of improving on the present bounds are extremely  
important.

\section*{Acknowledgements}
We acknowledge partial support from the following grants: 
PPARC Rolling Grants  PPA/G/ S2003/00096 and PP/D00036X/1; 
EU Network MRTN-CT-2004-503369; NATO grant PST.\ CLG.\ 980066; EU ILIAS RII3-CT-2004-506222. 

\appendix

\section{Zero global SUSY $\boldsymbol{F}$-terms and dummy fields\label{dummyfields}}

As we mentioned in the body of the text, a relatively simple solution to the
slight tension in the SUSY CP sector of the $SU(3)$ model under
consideration consists in arranging all the \textquotedblleft
global\textquotedblright\ SUSY $F$-terms in the visible sector ($W_{%
\overline{J}}^{\ast }$ in Eq.~(\ref{Ftermshape})) to vanish. This can be the
case if for instance we put $W_{\phi }=0$ by hand and (apart from demanding
that $\Sigma $ develops its GUT-scale breaking VEV in a SUSY-flat direction)
assume the family symmetry breakdown is triggered by $D$-terms, see e.g. \cite%
{deMedeirosVarzielas:2006fc}. In such a case, we get $x_{A}\approx 1$ for
all the flavons and also $x_{\Sigma }\approx 1$. If, on top of that, we
employ a \textquotedblleft dummy\textquotedblright\ $\Sigma _{0}$ field to
balance the powers of the first two operators in $W_{Y}$ so that they are
also dimension 7 as the last one with $\Sigma $, i.e.: 
\begin{eqnarray*}
\hat{Y}_{ab}^{f} &=&\frac{1}{M_{f}^{2}}[\frac{y_{1}^{f}}{M_{f}^{\Sigma }}%
(\phi _{123})_{a}(\phi _{23})_{b}\Sigma _{0}+\frac{y_{2}^{f}}{M_{f}^{\Sigma }%
}(\phi _{23})_{a}(\phi _{123})_{b}\Sigma _{0}+y_{3}^{f}(\phi _{3})_{a}(\phi
_{3})_{b} \\
&+&\frac{y_{\Sigma }^{f}}{M_{f}^{\Sigma }}(\phi _{23})_{a}(\phi
_{23})_{b}\Sigma ]
\end{eqnarray*}%
which, by the way, leads to the modified form of the trilinear dictionary: 
\begin{eqnarray*}
& & a_{1}^{f}=y_{1}^{f}\left( x_{123}+x_{23}+x_{\Sigma _{0}}\right)
,\;\; a_{\Sigma }^{f} =y_{\Sigma }^{f}\left( 2x_{23}+x_{\Sigma }\right),\; \\
& & a_{2}^{f}=y_{2}^{f}\left( x_{123}+x_{23}+x_{\Sigma _{0}}\right)
,\;\;A_{0}=m_{3/2}\;.
\end{eqnarray*}%
The critical bracket in Eq.~(\ref{bracket}) changes to: 
\begin{equation}
\left\vert x_{123}-x_{23}+x_{\Sigma ^{0}}-x_{\Sigma }\right\vert 
\label{bracketdummy}
\end{equation}%
and $x_{123,23}\approx x_{\Sigma ,\Sigma ^{0}}\approx 1$ again provides the
desired suppression in $\mu \rightarrow e\gamma $ and also in the relevant
EDMs.


\section{Dynamical suppression of flavon $\boldsymbol{F}$-terms \& $\boldsymbol{D}$-term cancellation 
\label{app2}}

It is straightforward to arrange a dynamical suppression of $x_{A}$ and/or 
$x_{\Sigma} $ in the $SU(3)$ model under consideration. 
To suppress all consider the superpotential of the form
\begin{equation}
W_{vis}\ni \lambda _{\Sigma }Z_{\Sigma }(\Sigma ^{2}-M_{\Sigma
}^{2})+\sum_{\Phi }\lambda _{\Phi }Z_{\Phi }(\Phi \overline{\Phi }-M_{\Phi
}^{2})+\ldots  \;.\label{fullW}
\end{equation}%
The discussion given in Section \ref{genericmechanism} applies with the
identification $\Psi =\overline{\Psi }\equiv \Sigma $ for the
Georgi-Jarlskog field $\Sigma $ (living in the adjoint of the underlying
Pati-Salam group) and similarly to the choice of $\Psi \equiv \Phi $ and $%
\overline{\Psi }\equiv \overline{\Phi }$ for each of the flavons of the $%
SU(3)$ model under consideration.

In a specific model like this one can easily inspect the effects of the $D$%
-terms that we have just touched upon in the preceding parts. It is clear
from Eq.~(\ref{fullW}) that in order for $\Sigma $ to admit for such a
quadratic term in the superpotential it must belong to a real representation
of the underlying Grand Unified group and thus its vevs must be real (up to
perhaps an irrelevant overall phase). The antisymmetry of the generators in
the real and unitary representation together with the canonicity of the
relevant K\"{a}hler metric then ensure vanishing of the corresponding $D$%
-term associated to $\Sigma $. The addition of the second term in Eq.~(\ref%
{fullW}) for each of the flavon species $\Phi $ in the model not only does
not disturb the $F_{\Sigma }$-suppression mechanism described above but
leads to the same suppression mechanism for each of the flavon $F$-terms.

Let us remark that the symbol $\overline{\Phi }$ in Eq.~(\ref%
{fullW}) denotes an additional conjugate flavon field, so that $%
\Phi \overline{\Phi }$ is a singlet under all the symmetries of the model.
Note that such extra flavons are usually needed anyway in order to cancel
the unwanted $D$-terms potentially arising at the $SU(3)$ family symmetry
breaking scale. At the level of the effective $SU(3)$ SUSY model \cite{Antusch:2007re} 
this is usually ensured by aligning manually the phases of
vevs of $\Phi $ and $\overline{\Phi }$ against against each other. This
follows immediately since the $F$-term has the form given in Eq.~(\ref{Fterms}%
) for each of the $SU(3)$ components and is minimized for $\langle\Phi\rangle _{a}=%
\langle\overline{\Phi }\rangle_{a.}$ This, in turn, ensures that the contribution of the
flavon fields to the $SU(3)$ $D$-terms vanishes (yielding $D_{a}^{\Phi
}=\Phi ^{\dagger }T_{a}\Phi $ and $D_{a}^{\overline{\Phi }}=\overline{\Phi }%
^{\dagger }\overline{T}_{a}\overline{\Phi }=-\overline{\Phi }^{\dagger
}T_{a}^{\ast }\overline{\Phi }$). 

To conclude, the dynamics of the system
under consideration (c.f.\ Eq.~(\ref{fullW})) leads to a natural
suppression of both $\Sigma $ and flavon $F$-terms and thus all the $x$%
-factors entering Eq.~(\ref{bracket}) can be made small, as desired. 
By restricting the superpotential to have  
a subset of the terms given in Eq.~(\ref{fullW}) it is straightforward to  
suppress only a subset of the F-terms.


\begin{thebibliography}{99}
\bibitem{Chung:2003fi} D.~J.~H. Chung \emph{et~al.}, \newblock Phys. Rept. 
\textbf{407}, 1 (2005), hep-ph/0312378.

\bibitem{King:2001uz} S.~F. King and G.~G. Ross, \newblock Phys. Lett. 
\textbf{B520}, 243 (2001), hep-ph/0108112.

\bibitem{King:2003rf} S.~F. King and G.~G. Ross, \newblock Phys. Lett. 
\textbf{B574}, 239 (2003), hep-ph/0307190.

\bibitem{Minkowski:1977sc} P.~Minkowski, \newblock Phys. Lett. \textbf{B67},
421 (1977).

\bibitem{King:1998jw} S.~F. King, \newblock Phys. Lett. \textbf{B439}, 350
(1998), hep-ph/9806440.

\bibitem{King:1999mb} S.~F. King, \newblock Nucl. Phys. \textbf{B576}, 85
(2000), hep-ph/9912492.

\bibitem{Harrison:2002er} P.~F. Harrison, D.~H. Perkins, and W.~G. Scott, %
\newblock Phys. Lett. \textbf{B530}, 167 (2002), hep-ph/0202074.

\bibitem{King:2005bj} S.~F. King, \newblock JHEP \textbf{08}, 105 (2005),
hep-ph/0506297.

\bibitem{deMedeirosVarzielas:2005ax} I.~de~Medeiros~Varzielas and G.~G.
Ross, \newblock Nucl. Phys. \textbf{B733}, 31 (2006), hep-ph/0507176.

\bibitem{Ross:2002mr} G.~G. Ross and O.~Vives, \newblock Phys. Rev. \textbf{%
D67}, 095013 (2003), hep-ph/0211279.

\bibitem{Ross:2004qn} G.~G. Ross, L.~Velasco-Sevilla, and O.~Vives, 
\newblock
Nucl. Phys. \textbf{B692}, 50 (2004), hep-ph/0401064.

\bibitem{Antusch:2007re}
  S.~Antusch, S.~F.~King and M.~Malinsky,
  JHEP {\bf 0806} (2008) 068 [arXiv:0708.1282 [hep-ph]].

\bibitem{Calibbi:2008qt} L.~Calibbi, J.~Jones-Perez and O.~Vives, 
arXiv:0804.4620 [hep-ph]. 


\bibitem{Georgi:1979df} H.~Georgi and C.~Jarlskog, 
Phys.\ Lett.\ B \textbf{86}, 297 (1979). 


\bibitem{Ross:2002fb} G.~G.~Ross and L.~Velasco-Sevilla, 
Nucl.\ Phys.\ B \textbf{653} (2003) 3 [arXiv:hep-ph/0208218]. 


\bibitem{deMedeirosVarzielas:2006fc} I.~de Medeiros Varzielas, S.~F.~King
and G.~G.~Ross, 
Phys.\ Lett.\ B \textbf{648} (2007) 201 [arXiv:hep-ph/0607045]. 


\bibitem{Ross:2007az} G.~Ross and M.~Serna, 
arXiv:0704.1248 [hep-ph]. 

\bibitem{Malinsky:2007pf} M.~Malinsky, \newblock (2007), arXiv:0710.2430
[hep-ph].

\bibitem{de Medeiros Varzielas:2006ma} I.~de Medeiros Varzielas and
G.~G.~Ross, 
arXiv:hep-ph/0612220. 


\end{thebibliography}
\end{document}